\documentclass[aps,prl,superscriptaddress,twocolumn,floatfix,showpacs,amsfonts,amssymb,amsmath]{revtex4-1}
\usepackage{graphicx}

\begin{document}
\title{Crossover from vibrational to rotational collectivity in heavy nuclei in the shell-model Monte Carlo approach}

\author{C.~\surname{\"{O}zen}}
\affiliation{Center for Theoretical Physics, Sloane Physics Laboratory,Yale University, New Haven, CT 06520, USA}
\affiliation{Faculty of Engineering and Natural Sciences, Kadir Has University, Istanbul 34083, Turkey}
\author{Y.~\surname{Alhassid}}
\affiliation{Center for Theoretical Physics, Sloane Physics Laboratory,Yale University, New Haven, CT 06520, USA}
\author{H.~\surname{Nakada}}
\affiliation{Department of Physics, Graduate School of Science,Chiba University, Inage, Chiba 263-8522, Japan}
\date{\today}

\begin{abstract}
Heavy nuclei exhibit a crossover from vibrational to rotational collectivity as the number of neutrons or protons increases from shell closure towards midshell, but the microscopic description of this crossover has been a major challenge. We apply the shell model Monte Carlo approach to families of even-even samarium and neodymium isotopes and identify a microscopic signature of the crossover from vibrational to rotational collectivity in the low-temperature behavior of $\langle \mathbf{J}^2 \rangle_T$, where $\bf J$ is the total spin and $T$ is the temperature. This signature agrees well with its values extracted from experimental data. We also calculate the state densities of these nuclei and find them to be in very good agreement with experimental data. Finally, we define a collective enhancement factor from the ratio of the total state density to the intrinsic state density as calculated in the finite-temperature Hartree-Fock-Bogoliubov approximation. The decay of this enhancement factor with excitation energy is found to correlate with the pairing and shape phase transitions in these nuclei.
\end{abstract}


\maketitle

{\it Introduction.}  For microscopic calculation of the statistical and collective properties of atomic nuclei, and in particular level densities, the shell model Monte Carlo (SMMC) method~\cite{LAN93,ALH94} has proved to be particularly useful~\cite{NA97,OR97,ALN99,ABLN00, ALN07,OZE07}. This method enables fully correlated configuration-interaction (CI) shell-model calculations in much larger configuration spaces than those that can be treated by conventional diagonalization methods. It has been extended to heavy nuclei and applied to the well-deformed rare-earth nucleus $^{162}$Dy by implementing a proton-neutron formalism and a stabilization technique in the canonical ensemble~\cite{ALH08}.

Here we use the SMMC method to study microscopically families of even-even rare-earth isotopes. Such isotopic families exhibit a crossover from vibrational to rotational collectivity as the number of neutrons increases from shell closure towards the midshell region. This crossover corresponds, in the thermodynamic limit, to a phase transition from spherical to deformed nuclei. The microscopic description of such a crossover in the framework of a truncated spherical shell model has proved challenging because the dimensionality of the many-particle shell model space required to describe heavy rare-earth nuclei is many orders of magnitude beyond the capability of conventional diagonalization methods. 

The SMMC approach, while capable of treating large model spaces, does not provide the detailed spectroscopic information that is often used to identify the type of nuclear collectivity. Here we show that the crossover from vibrational to rotational collectivity can be alternatively identified by the low-temperature behavior of $\langle \mathbf{J}^2 \rangle_T$, where $\bf J$ is the total nuclear spin and $T$ is the temperature. This thermal observable can be calculated in the SMMC method, and we use it to demonstrate that the vibrational-to-rotational crossover in families of even-even samarium and neodymium isotopes can be described microscopically in the framework of a truncated spherical shell model approach. Furthermore, we find that the temperature dependence of $\langle \mathbf{J}^2 \rangle_T$ agrees well with its values extracted from experimental data.  We also calculate the total state densities $\rho(E_x)$ for the corresponding samarium and neodymium isotopes and find them in very good agreement with experimental state densities.

Vibrational and rotational collective states account for a significant fraction of the total state density up to moderate excitation energies, and their contribution is described by the so-called collective enhancement factor. Collective enhancement is one of the least understood topics in the study of level densities~\cite{RIPL}. Both empirical and combinatorial models of level densities often use phenomenological enhancement factors~\cite{HAN83,IGN85}. Although various expressions for vibrational and rotational collective enhancement factors are available in the literature~\cite{RIPL,KON08}, it is highly desirable to study such enhancement factors microscopically. In particular, little is known about the decay of collectivity with excitation energy although it plays an important role in fission reactions~\cite{RIPL}. Here we define a total collective enhancement factor as the ratio between the total state density and the intrinsic state density obtained within the thermal Hartree-Fock-Bogoliubov (HFB) approximation and study microscopically the decay of this enhancement factor with excitation energy. We find that the damping of the vibrational and rotational collectivity with excitation energy is correlated, respectively, with the pairing and shape phase transitions in these nuclei.

{\it Model space and interaction.} Here we use the same single-particle model space as in Ref.~\cite{ALH08}, namely
$0g_{7/2}$, $1d_{5/2}$, $1d_{3/2}$, $2s_{1/2}$, $0h_{11/2}$, $1f_{7/2}$
for protons, and $0h_{11/2}$, $0h_{9/2}$, $1f_{7/2}$, $1f_{5/2}$,
$2p_{3/2}$, $2p_{1/2}$, $0i_{13/2}$, and $1g_{9/2}$ for neutrons.
This model space is larger than one major shell for both protons and neutrons, and was determined by examining the occupation probabilities of spherical orbitals for well-deformed rare-earth nuclei~\cite{ALH08}.

The single-particle energies in the CI shell-model Hamiltonian are determined so as to reproduce the single-particle energies of a spherical Woods-Saxon plus spin-orbit potential in the spherical Hartree-Fock approximation. The effective interaction consists of monopole pairing and multipole-multipole terms (quadrupole, octupole and hexadecupole)~\cite{ALH08}
\begin{equation}\label{interaction}
 - \!\! \sum_{\nu=p,n} g_\nu P^\dagger_\nu P_\nu
 - \!\! \sum_\lambda \chi_\lambda : (O_{\lambda;p} + O_{\lambda;n})\cdot
 (O_{\lambda;p} + O_{\lambda;n})\!: \;.
\end{equation}
Here $:\,:$ denotes normal ordering,
$P^\dagger_\nu = \sum_{nljm}(-)^{j+m+l}a^\dagger_{\alpha jm;\nu}
a^\dagger_{\alpha j-m;\nu}$ are monopole pair operators for protons  ($\nu=p$) and neutrons ($\nu=n$), while  $O_{\lambda;\nu}=\frac{1}{\sqrt{2\lambda+1}}
\sum_{ab}\langle j_a||\frac{dV_\mathrm{WS}}{dr}Y_\lambda||j_b\rangle
[a^\dagger_{\alpha j_a;\nu}\times \tilde{a}_{\alpha j_b;\nu}]^{(\lambda)}$
with $\tilde{a}_{jm}=(-)^{j+m}a_{j-m}$ is the $2^\lambda$-pole operator. The pairing coupling strengths are parametrized by  $g_\nu=\gamma\cdot \bar{g}_\nu$
with $\bar{g}_p=10.9/Z$ and $\bar{g}_n=10.9/N$ ($Z$ and $N$ are the number of protons and neutrons, respectively). The latter are determined so that the pairing gaps calculated in the number-projected BCS approximation could reproduce the experimental even-odd mass differences for spherical nuclei in the mass region~\cite{ALH08}.
The factor $\gamma$ is an effective suppression factor of the overall pairing strength, part of which may be ascribed to the fluctuations induced by pairing correlations beyond the number-projected BCS approximation. The multipole-multipole interaction terms we include in (\ref{interaction}) are the quadrupole, octupole and hexadecupole terms (i.e., $\lambda=2,3,4$). Their strengths are given by $\chi_\lambda=\chi\cdot k_\lambda$, where $\chi$ is determined self-consistently~\cite{abdk96}
and $k_\lambda$ are renormalization factors accounting for core polarization effects.

In general, the moment of inertia $\mathcal{I}$ of the ground-state band for a deformed nucleus is sensitive to $\gamma$, while the slope of $\ln \rho(E_x)$ is sensitive to $k_2$~\cite{ALH08}. In Ref.~\cite{ALH08} we have adopted the values $\gamma=0.77$, $k_2=2.12$, $k_3=1.5$ and $k_4=1$ for $^{162}$Dy.  We have studied families of samarium ($^{148-155}$Sm) and neodymium ($^{143-152}$Nd) isotopes (both even and odd) and found that a more appropriate choice to reproduce the overall experimental systematics is $k_3=1$, while $\gamma$ and $k_2$ are parametrized by a weak and smooth $N$-dependence $\gamma = 0.72- 0.5/[(N-90)^2+5.3]$ and $k_2 = 2.15+0.0025(N-87)^2$.

\begin{figure}[b]
    \includegraphics[clip,angle=0,width=0.8\columnwidth]{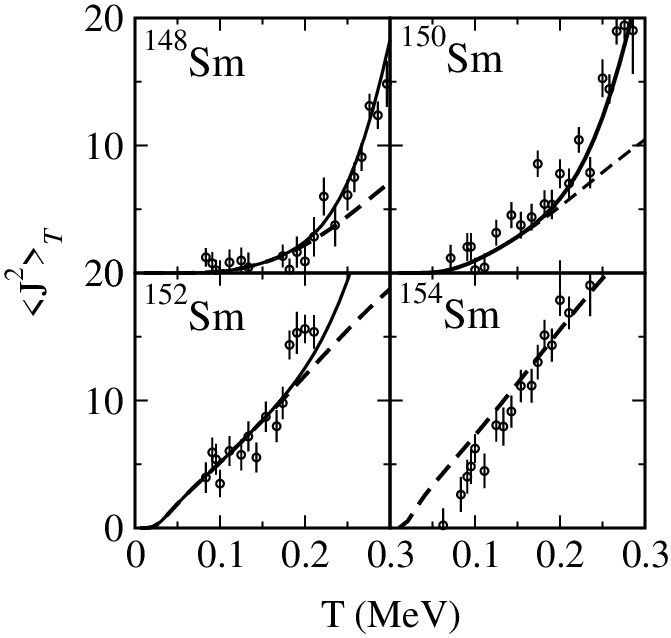}
    \caption{$\langle \mathbf{J}^2 \rangle_T$ as a function of temperature in a family of even-even samarium isotopes $^{148-154}$Sm. The SMMC results (open circles) are
    compared with the experimental results deduced from known low-lying levels (dashed lines) and
    from the additional contribution of higher levels described by an experimental BBF level density (solid lines).}
    \label{Fig:Sm-J2}
\end{figure}

\textit{The crossover from vibrational to rotational collectivity.}
At low temperatures, the observable $\langle \mathbf{J}^2 \rangle_T$ is dominated by the ground-state band. Assuming a vibrational or rotational ground-state band with an excitation energy $E_{2^+} $ of the first excited $J=2^+$ state, we find~\cite{ALH08,FAN05}
\begin{eqnarray}\label{J2-theory}
\langle \mathbf{J}^2 \rangle_T \approx
 \left\{ \begin{array}{cc}
 30  \frac{e^{-E_{2^+}/T}}{\left(1-e^{- E_{2^+}/T}\right)^2} &{\rm vibrational\; band}  \\
 \frac{6}{E_{2^+}} T & {\rm rotational \;band}
 \end{array} \right. \;.
\end{eqnarray}
Thus, the low-temperature behavior of $\langle \mathbf{J}^2 \rangle_T$ is sensitive to the type of collectivity and can be used to distinguish between vibrational and rotational nuclei.

In Fig.~\ref{Fig:Sm-J2}, we show the SMMC results (open circles) for $\langle \mathbf{J}^2 \rangle_T$ at low temperatures for the even-even samarium isotopes $^{148-154}$Sm. The $^{148}$Sm nucleus exhibits a soft response to temperature, typical of a vibrational nucleus. Indeed, the vibrational band formula in Eq.~(\ref{J2-theory}) can be well fitted to the SMMC results for $\langle \mathbf{J}^2 \rangle_T$ with $E_{2^+}^{\mathrm{vib}}=0.538 \pm 0.031$ MeV, in agreement with the experimental value of $E_{2^+}^{\mathrm{exp}}=0.550$ MeV.
In the heavier samarium isotopes, the low-temperature response of $\langle \mathbf{J}^2 \rangle_T$ becomes increasingly linear, suggesting
the presence of stronger rotational collectivity. Fitting the rotational band formula in  Eq.~(\ref{J2-theory}) to the SMMC results for $^{154}$Sm, we find  $E_{2^+}^{\mathrm{rot}}=0.087 \pm 0.006$ MeV, consistent with the experimental value of  $E_{2^+}^{\mathrm{exp}}=0.082$ MeV -- an evidence for the rotational nature of this nucleus. Thus our SMMC results for $\langle \mathbf{J}^2 \rangle_T$ reproduces the proper dominant collectivity in both $^{148}$Sm and ${}^{154}$Sm, demonstrating the crossover from vibrational to rotational collectivity in the even-even isotopic chain.

The experimental values of $\langle \mathbf{J}^2 \rangle_T$ can be extracted at sufficiently low temperatures from
\begin{equation}
 \label{Eq:J2low}
 \langle \mathbf{J}^2 \rangle_T = \frac{\sum_i J_i(J_i+1)(2J_i+1)e^{-E_{i}/T}}{\sum_i (2J_i+1)e^{-E_{i}/T}}\;,
\end{equation}
where the summations are carried over the experimentally known low-lying energy levels $i$ with excitation energy $E_i$  and spin $J_i$. These experimental estimates are shown by the dashed lines in Fig.~\ref{Fig:Sm-J2} for the even-even  $^{148-154}$Sm isotopes.  However, since the experimental level scheme is incomplete above a certain energy, Eq.~(\ref{Eq:J2low}) underestimates the  experimental value of $\langle \mathbf{J}^2 \rangle_T$ above a certain temperature. We can obtain a more realistic estimate by using the discrete sum over energy levels up to a certain energy threshold $E_N$ (below which the experimental spectrum is complete), and estimate the contribution of levels above $E_N$ in terms of an average state density $\rho(E_x)$  that is parametrized  with the help of available experimental data.  We then have
\begin{eqnarray}
\label{Eq:J2high}
 \langle \mathbf{J}^2 \rangle_T & = & \frac{1}{Z(T)} \left(\sum_i^N J_i(J_i+1)(2J_i+1)e^{-E_{i}/T} +  \right. \nonumber \\
   & & \left. \int_{E_{N}}^\infty d E_x \: \rho(E_x) \: \langle \mathbf{J}^2 \rangle_{E_x} \; e^{-E_x/T} \right)\;,
\end{eqnarray}
with $Z(T)=\sum_{i}^{N} (2J_i+1) e^{-E_i/T} + \int_{E_{N}}^\infty d E_x \rho(E_x) e^{-E_x/T}$ is the corresponding experimental partition function. Here $\langle \mathbf{J}^2 \rangle_{E_x}$ is the average value of $\mathbf{J}^2$ at a given excitation energy $E_x$. For the level density we use a backshifted Bethe formula (BBF) with single-particle level density parameter $a$ and backshift parameter $\Delta$, extracted from the neutron resonance data (when available) and counting data at low energies. Using the spin-cutoff model (obtained assuming random coupling of the individual nucleon spins~\cite{ER60}), we have $\langle \mathbf{J}^2 \rangle_{E_x} = 3\langle \mathrm{J}^2_z \rangle_{E_x}=3\sigma^2(E_x)$
where $\sigma^2$ is the spin-cutoff parameter. The latter is estimated from $\sigma^2=\mathcal{I}T/\hbar^2$ using $T=[(E_x-\Delta)/a]^{1/2}$  and a rigid-body moment of inertia $\mathcal{I}\approx 0.015 A^{5/3}\hbar^2$.  The corresponding results for $\langle \mathbf{J}^2 \rangle_T$, shown by the solid lines in Figs.~\ref{Fig:Sm-J2}, are in reasonable agreement with the SMMC results along the crossover from  $^{148}$Sm to $^{154}$Sm.

\begin{figure}[t]
    \includegraphics[clip,angle=0,width=0.8\columnwidth]{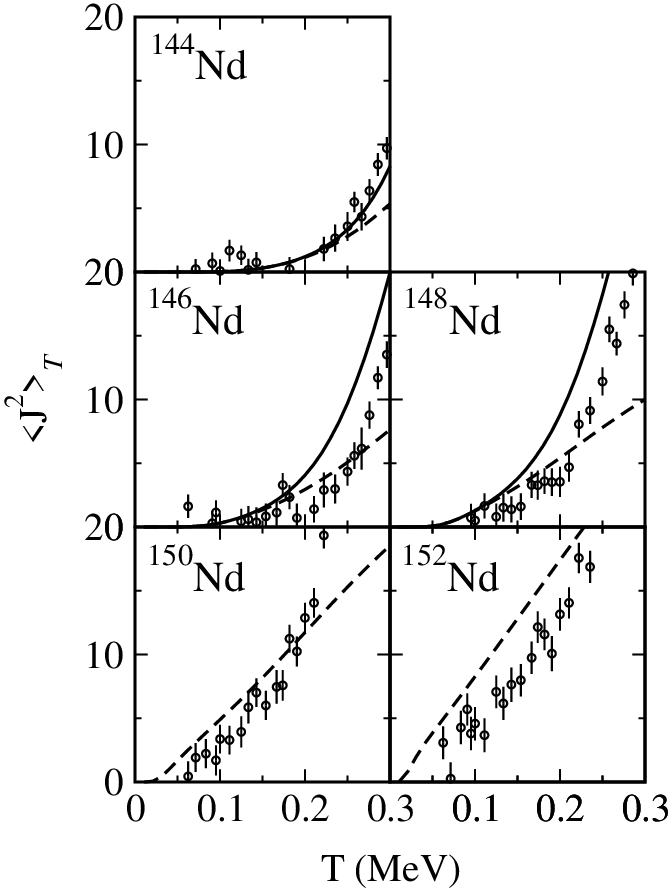}
    \caption{$\langle \mathbf{J}^2 \rangle_T$ as a function of temperature in a family of even-even neodymium isotopes $^{144-152}$Nd. Symbols and lines are as in
    Fig.~\ref{Fig:Sm-J2}.}
    \label{Fig:Nd-J2}
\end{figure}

In Fig.~\ref{Fig:Nd-J2} we show similar results for the low-temperature behavior of $\langle \mathbf{J}^2 \rangle_T$ for the even-even $^{144-152}$Nd isotopes. The  $\langle \mathbf{J}^2 \rangle_T$ response at low temperatures---soft in $^{144}$Nd--- becomes more rigid in the heavier neodymium isotopes to assume an approximately linear form in $^{150}$Nd and $^{152}$Nd.
Fitting the SMMC results to the vibrational band formula in Eq.~(\ref{J2-theory}) for $^{144}$Nd  we find  $E_{2^+}^{\mathrm{vib}}=0.702 \pm 0.062$ MeV, in agreement with the experimental value of $E_{2^+}^{\mathrm{exp}}=0.697$ MeV. Using the rotational band formula, we find $E_{2^+}^{\mathrm{rot}}=0.132 \pm 0.012$ MeV for $^{150}$Nd ($E_{2^+}^{\mathrm{exp}}=0.130$ MeV) and  $E_{2^+}^{\mathrm{rot}}=0.107 \pm 0.006$ MeV for ${}^{152}$Nd  ($E_{2^+}^{\mathrm{exp}}=0.073$ MeV).
 These results confirm that our spherical shell model Hamiltonian is capable of describing the crossover from vibrational collectivity in $^{144}$Nd to rotational collectivity in $^{150}$Nd and $^{152}$Nd.

\begin{figure}[tbh]
\includegraphics[clip,angle=0,width=0.8\columnwidth]{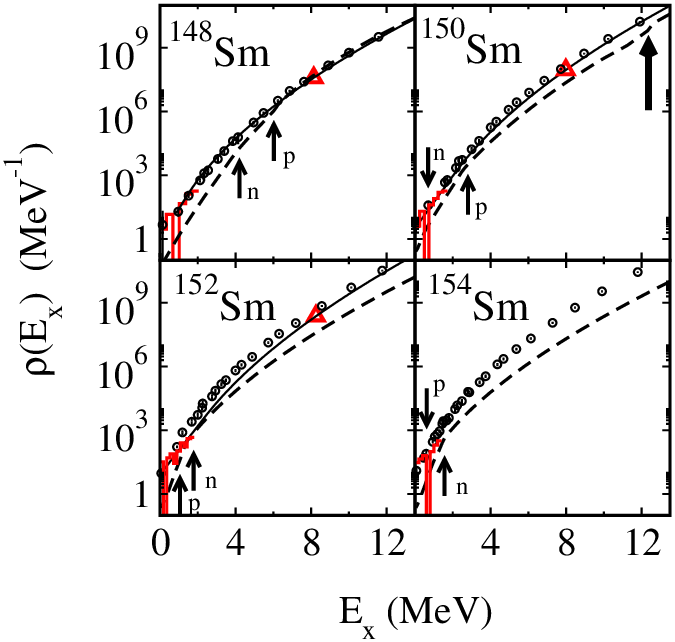}
\caption{Total state densities in the even-even $^{148-154}$Sm isotopes. The SMMC results (open circles) are compared with level counting data (histograms), neutron resonance data (triangles) and the BBF parametrization of the experimental data (solid lines). Also shown are the HFB level densities (dashed lines). The neutron and proton pairing transitions are indicated by arrows and the shape transition by a thick arrow.}
\label{Fig:Sm-rho}
\end{figure}

\textit{The determination of the ground-state energy for even-even isotopes.}
An accurate estimate of the ground-state energy $E_0$ is crucial in obtaining the excitation energy $E_x=E-E_0$ necessary for the calculation of the state density. Because of the low excitation energies in the heavy rare-earth nuclei, we have carried out calculation of the thermal energy up to an inverse temperature value of $\beta\,(=1/T)\sim 20$ MeV$^{-1}$~\cite{ALH08}.
The ground-state energy can then be determined as follows. In vibrational and rotational nuclei we have used expressions for the low-temperature energy in the ground-state band approximation~\cite{ALH08,FAN05}
\begin{eqnarray}\label{E-theory}
E(T) \approx
 \left\{ \begin{array}{cc}
 E_0 + 5 E_{2^+} {\frac{e^{-E_{2^+}/T}}{1-e^{- E_{2^+}/T}} }  &{\rm vibrational\; band}  \\
 E_0 + T & {\rm rotational \;band}
 \end{array} \right.
\end{eqnarray}
to extract the ground-state energy $E_0$. For other nuclei in the crossover we have estimated $E_0$ by taking an average value of  $E(T)$ at sufficiently low temperatures.

\textit{State densities: theory and experiment.}
In Fig.~\ref{Fig:Sm-rho} we show the total state densities as a function of the excitation energy $E_x$ for the even-even samarium isotopes ($^{148-154}$Sm). The SMMC state densities (circles), calculated using the methods of Refs.~\cite{NA97} and \cite{ALH08}, are compared with experimental data that consist of level counting data at low excitation energies (histograms) and, when available, neutron resonance data at the neutron threshold energy (triangles). For nuclei with neutron resonance data, we have also included a BBF state density~\cite{DILG73} (solid lines) whose parameters $a$ and $\Delta$ are determined from the level counting and the neutron resonance data~\cite{OZE12}. For the SMMC state densities of the even-even $^{144-152}$Nd isotopes (not shown) we find similar agreement with experimental data.

For comparison, we also show in Fig.~\ref{Fig:Sm-rho} the level density $\rho_{\rm HFB}$ calculated from the finite-temperature HFB approximation (dashed lines) using the same Hamiltonian. The HFB level density accounts only for intrinsic states, and the enhancement observed in the SMMC state density originates in rotational bands that are built on top of these intrinsic states as well as in vibrational collectivity that is missed in the HFB approximation. The kinks in $\rho_{\rm HFB}$ are associated with the proton and neutron pairing phase transitions (arrows) and the shape phase transition (thick arrow). $^{148}$Sm is spherical in its ground state and undergoes pairing transitions only.  $^{150}$Sm has a non-zero deformation in its ground state and undergoes also a shape transition to a spherical shape at $E_x \approx 12.5$ MeV. The ground-state deformation continues to increase with mass number in  $^{152}$Sm and  $^{154}$Sm,  and the shape transitions occur at higher excitation energies (outside the energy range shown in the figure).

\begin{figure}[t!]
    \includegraphics[clip,angle=0,width=0.8\columnwidth]{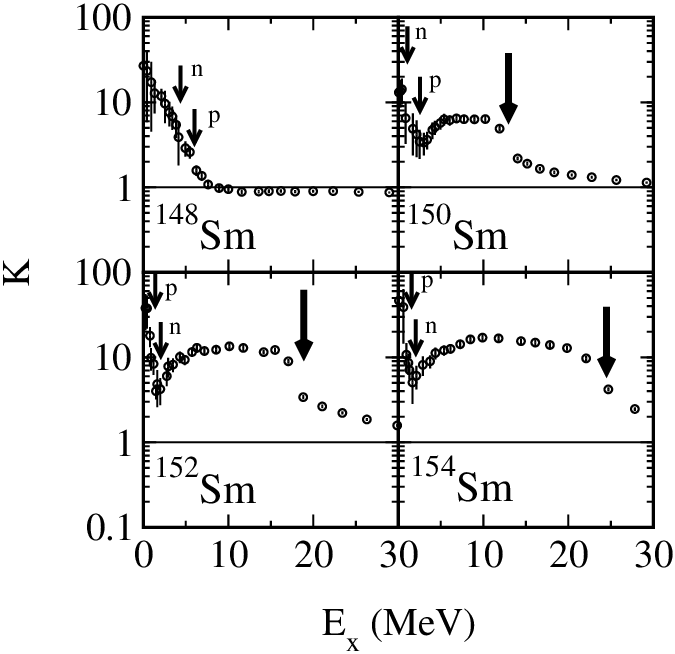}
    \caption{Total collective enhancement factor $K$ (see text) in the even-even $^{148-154}$Sm isotopes as a function of excitation energy $E_x$. Arrows are as in Fig.~\ref{Fig:Sm-rho}.
     }
    \label{Fig:Sm-enh}
\end{figure}

\textit{Collective enhancement factors.}  The enhancement of level densities due to collective effects is difficult to calculate microscopically and is often modeled by phenomenological formulas. Here we propose to define a collective enhancement factor by the ratio $K = \rho_{\rm SMMC}/\rho_{\rm HFB}$, a quantity that we can extract directly in our microscopic CI shell model approach. In Fig.~\ref{Fig:Sm-enh}, we show $K$ (on a logarithmic scale) versus excitation energy $E_x$ for the same samarium isotopes of Fig.~\ref{Fig:Sm-rho} but up to higher excitation energies of $E_x \sim 30$ MeV.

$^{148}$Sm is spherical in its ground state and the observed collective enhancement must be due to vibrational collectivity. This collectivity disappears (i.e, $K\approx 1$) above the proton pairing transition. The other samarium isotopes shown in Fig.~\ref{Fig:Sm-enh} are deformed in their ground state and $K$ exhibits a local minimum above the pairing transitions, which we interpret as the decay of vibrational collectivity. The rapid increase of $K$ above the pairing transitions originates in rotational collectivity. This collectivity reaches a plateau as a function of excitation energy and then decay gradually to $K\sim 1$ in the vicinity of the shape transition (thick arrow) when the nucleus becomes spherical and no longer supports rotational bands.

\textit{Conclusions.} We have carried out SMMC calculations for isotopic families of the even-even rare-earth nuclei ${}^{148-154}$Sm and ${}^{144-152}$Nd. Using the observable $\langle \mathbf{J}^2 \rangle_T$, whose low-temperature behavior is sensitive to the specific type of nuclear collectivity, we have demonstrated that a truncated spherical shell model approach can describe the crossover from vibrational to rotational collectivity in heavy nuclei.  We
have also calculated the total SMMC state densities and found them to be in very good agreement with experimental data. We have extracted microscopically a collective enhancement factor defined as the ratio between the SMMC and HFB state densities. The damping of vibrational and rotational collectivity seems to  correlate with the pairing and shape phase transitions, respectively.

This work was supported in part by the U.S. DOE grant No. DE-FG02-91ER40608, and by the JSPS Grant-in-Aid for Scientific Research (C) No.~22540266. Computational cycles were provided by the NERSC high performance computing facility at LBL and by the facilities of the Yale University Faculty of Arts and Sciences High Performance Computing Center.


\begin{thebibliography}{99}

\bibitem{LAN93} G.H.~Lang, C.W.~Johnson, S.E.~Koonin, and W.E.~Ormand,
Phys. Rev. C \textbf{48}, 1518 (1993).
\bibitem{ALH94} Y.~Alhassid, D.J.~Dean, S.E.~Koonin, G.~Lang, and W.E.~Ormand,
Phys. Rev. Lett. \textbf{72}, 613 (1994).
\bibitem{NA97} H. Nakada and Y. Alhassid, Phys. Rev. Lett. {\bf 79}, 2939 (1997).
\bibitem{OR97} W.E. Ormand, Phys. Rev. C {\bf 56}, R 1678 (1997).
\bibitem{ALN99} Y. Alhassid, S. Liu and H. Nakada, Phys. Rev. Lett.
{\bf 83}, 4265 (1999).
\bibitem{ABLN00} Y. Alhassid, G.F. Bertsch, S. Liu and H. Nakada,
\emph{Phys. Rev. Lett.} \textbf{84}, 4313 (2000).
\bibitem{ALN07} Y. Alhassid, S. Liu, and H. Nakada, Phys. Rev. Lett. {\bf 99}, 162504 (2007).
\bibitem{OZE07} C.~\"{O}zen, K.~Langanke, G.~Martinez-Pinedo, and D.~J.~Dean, Phys. Rev. C {\bf 75}, 064307 (2007).
\bibitem{ALH08} Y. Alhassid, L. Fang and H. Nakada, Phys. Rev. Lett. {\bf 101}, 082501 (2008).
\bibitem{RIPL} R. Capote et. al., Nuclear Data Sheets {\bf 110}, 3107 (2009), and references therein.
\bibitem{HAN83} G.~Hansen and A.S.~Jensen, Nucl. Phys. A {\bf 406}, 236 (1983).
\bibitem{IGN85} A.V. Ignatyuk, IAEA Report No.~INDC(CCP)-233, 1985.
\bibitem{KON08} A.J. Koning, S.Hilaire and S. Goriely, Nuc. Phys. A {\bf 810}, 13 (2008).
\bibitem{abdk96} Y. Alhassid, G.F. Bertsch, D.J. Dean and S.E. Koonin,
 Phys. Rev. Lett. {\bf 77}, 1444 (1996).
\bibitem{FAN05} L. Fang, Ph.D. thesis, Yale University (2005).
\bibitem{ER60} T. Ericson, Adv. Phys. {\bf 9}, 425 (1960).
\bibitem{DILG73} W.~Dilg, W.~Schantl, H.~Vonach, M.~Uhl, Nucl.~Phys.~A {\bf 217}, 269 (1973).
\bibitem{OZE12} C.~\"{O}zen and Y.~Alhassid (to be published).

\end{thebibliography}
\end{document}